\documentclass[12pt]{article}
\usepackage{amssymb,amsmath}
\usepackage[noblocks]{authblk}
\usepackage[top=0.75in, bottom=0.75in, left=0.75in, right=0.75in, dvips]{geometry}
\usepackage{caption}
\date{}

\begin{document}
\textwidth 10.0in   
\textheight 9.0in 
\topmargin -0.60in
\title{Derivation of Gauge Symmetries in Supergravity\\
with a Cosmological Constant in $2 + 1$ Dimensions}
\author[1,2]{D.G.C. McKeon}
\affil[1] {Department of Applied Mathematics, The
University of Western Ontario, London, ON N6A 5B7, Canada} 
\affil[2] {Department of Mathematics and
Computer Science, Algoma University, 
Sault Ste. Marie, ON P6A 2G4, Canada}
\date{}
\maketitle

\maketitle
\noindent
email: dgmckeo2@uwo.ca

\begin{abstract}
The canonical structure of supergravity with a cosmological constant is analyzed in $2 + 1$ dimensions using the Dirac constraint formalism.  Using the approach of Henneaux, Teitelboim and Zanelli, the first class constraints are used to find the local gauge symmetries of this model.  Provided the cosmological constant is negative, this novel gauge algebra closes, without having to invoke the equations of motion or introducing auxiliary fields.  There are two Bosonic and one Fermionic gauge symmetries.
\end{abstract}

We will consider a model defined by the action\footnote{We use the notation of ref. [4] as reiterated in the appendix.}
\begin{align}
S = \int d^3x \epsilon^{\mu \nu \lambda} \Bigg[ b_\mu^i R_{\nu\lambda i} &+ \overline{\psi}_\mu D_\nu \psi_\lambda + \frac{\Lambda}{3} \epsilon_{ijk} b_\mu^i b_\nu^j b_\lambda^k \nonumber \\
&+ \frac{iK}{2} \overline{\psi}_\mu b_\nu^i \gamma_i \psi_\lambda\Bigg] .
\end{align}

The action of eq. (1) is a first order action with $b_\mu^i$ and $\omega_\mu^i$ being independent fields.  The terms proportional to $\Lambda$ and $K$ are introduced independently.  Consequently this action distinct from the $2 + 1$ dimensional supergravity actions considered in refs. [1,2].  In eq. (1), $\Lambda$ is the cosmological constant; the closure of the gauge algebra will fix $\Lambda$ to be negative with the novel relation
\begin{equation}
\Lambda = -K^2.
\end{equation}
Interesting black hole solutions occur in $2 + 1$ dimensional space-times in which there is a negative cosmological constant [3].

The momenta conjugate to $b_0^i \equiv b^i$, $\omega^i_0 \equiv \omega^i$ and $\psi_0 \equiv \psi$ all vanish, giving respectively the primary constraints
\begin{equation}\tag{3a,b,c}
p_i = I\!\!P_i = \pi = 0.
\end{equation}

Additional primary constraints are the momenta conjugate to $b^{\alpha i}$, $\omega^{\alpha i}$ and $\psi^\alpha$
\begin{equation}\tag{4a}
p_{\alpha i} = 0
\end{equation}
\begin{equation}\tag{4b}
I\!\!P_{\alpha i} - 2 \epsilon_{\alpha\beta}b_i^\beta = 0
\end{equation}
\begin{equation}\tag{4c}
\pi_\alpha + \epsilon_{\alpha\beta} \overline{\psi}^\beta = 0.
\end{equation}
Defining the canonical Hamiltonian to be
\begin{equation}\tag{5}
H_c = \dot{q}_i p_i + \dot{\psi}_i \pi_i - L
\end{equation}
with the primary constraints arising from $p_i = \frac{\partial L}{\partial \dot{q}_i}, \pi_i = \frac{\partial L}{\partial \dot{\psi}_i}$ being used, we find that
\begin{align}
H_c &= -b^i \left[ \epsilon^{\alpha\beta}\left( R_{\alpha\beta i} + \Lambda\epsilon_{ijk} b_\alpha^j b_\beta^k - \frac{iK}{2} \overline{\psi}_\alpha \gamma_i \psi_\beta \right)\right]\nonumber \\
&- 2\overline{\psi} \left[ \epsilon^{\alpha\beta}\left( D_\alpha \psi_\beta  + \frac{iK}{2} b^i_\alpha \gamma_i \psi_\beta \right)\right]\nonumber \\
&-2\omega^i \left[ \epsilon^{\alpha\beta}\left( \partial_\alpha  b_{i\beta} - \epsilon_{ijk} 
\omega_\alpha^j b_\beta^k - \frac{i}{4} \overline{\psi}_\alpha 
\gamma_i \psi_\beta\right)\right]\nonumber \\
&\equiv -b^i \Phi_{1i} - 2 \overline{\psi}\Psi - 2\omega^i \Phi_{2i}.\tag{6}
\end{align}
The constraints of eq. (4) are second class; they result in the Dirac brackets [4]
\begin{align}
\left\lbrace b^i_\alpha, \omega_\beta^j \right\rbrace^* &= \frac{1}{2}\eta^{ij}\epsilon_{\alpha\beta}
\tag{7a}\\
\intertext{and}
\left\lbrace \psi_\alpha , \overline{\psi}_\beta \right\rbrace^* &= \frac{1}{2}\epsilon_{\alpha\beta}.\tag{7b}
\end{align}
With the Dirac brackets of eq. (7) we find that the primary constraints of eq. (3) are first class and that they result in the secondary first class constraints $\Phi_{1i}$, $\Phi_{2i}$, $\Psi$ whose algebra is 
\begin{align}
\left\lbrace \Phi_{1i}, \Phi_{1j} \right\rbrace^* &= - 2K^2 \epsilon_{ijk}\Phi_2^k \tag{8a}\\
\left\lbrace \Phi_{2i}, \Phi_{2j} \right\rbrace^* &= - \frac{1}{2} \epsilon_{ijk}\Phi_2^k \tag{8b}\\
\left\lbrace \Phi_{1i}, \Phi_{2j} \right\rbrace^* &= - \frac{1}{2} \epsilon_{ijk}\Phi_1^k \tag{8c}\\
\left\lbrace \Psi, \overline{\Psi} \right\rbrace^* = & \left(- \frac{i}{8} \Phi_{1i} - \frac{iK}{4}\Phi_{2i}\right)\gamma^i \tag{8d}\\
\left\lbrace \Psi, \Phi_{1i}\right\rbrace^* = & \frac{iK}{2}\gamma_i \Psi \tag{8e}\\
\left\lbrace \Psi, \Phi_{2i}\right\rbrace^* = & \frac{i}{4}\gamma_i \Psi \tag{8f}
\end{align}
provided eq. (2) is satisfied.  It can be verified that the Jacobi identities are consistent with this algebra.  If $K = 0$, we recover the constraint algebra of ref. [4].

In eq. (1), there are 18 Bosonic fields ($b_\mu^i, \omega_\mu^i$) so in phase space there are 36 Bosonic fields (including the canonical momenta $p_i^\mu, I\!\!P_i^\mu$).  There are also six primary first class constraints (eqs. (3a,b)), twelve primary second class constraints (eqs. (4a,b)), six secondary first class constraints ($\Phi_{1i}, \Phi_{2i}$).  Each of the twelve first class constraints requires an associated gauge condition.  Since each constraint and gauge condition eliminates a degree of freedom in phase space, we see that there are no Bosonic physical degrees of freedom left in eq. (1).  Similarly, the six independent Fermionic fields in the Majorana spinor $\psi_\mu$ are all non physical as there are two primary first class constraints (eq. (3c)), four primary second class constraints (eq. (4c)), two secondary first class constraints ($\Psi$), as well as a gauge condition associated with each first class constraint.

Suppose that, upon using Dirac brackets to eliminate the second class constraints in a model, the remaining dynamical variables consist of two sets of canonically conjugate pairs ($Q_i, I\!\!P_i$) and ($q_i, p_i$) so that the canonical Hamiltonian is of the form
\begin{equation}\tag{9}
H_c = -Q_i \Phi_i(q_j, p_j).
\end{equation}
We also assume that
\begin{equation}\tag{10}
I\!\!P_i = 0
\end{equation}
is a set of primary first class constraints and $\Phi_i(q_j, p_j)$ is a set of secondary first class constraints satisfying
\begin{equation}\tag{11}
\left\lbrace \Phi_i, \Phi_j \right\rbrace^* = c_{ijk} \Phi_k.
\end{equation}

In ref. [5], Henneaux, Teitelboim and Zanelli (HTZ) outline how first class constraints can be used to find the gauge transformations that leave an action invariant.\footnote{There may be additional gauge transformations that do not follow from the first class constraints; such is the case with the Palatini form of the Einstein-Hilbert action in $1 + 1$ dimensions [7].} (An alternate approach is due to Castellani [6].)  In this approach, a gauge generator $G$ is a linear combination of the first class constraints so that in this case
\begin{equation}\tag{12}
G = \lambda_i I\!\!P_i + \Lambda_i \Phi_i \equiv \mu_i \gamma_i\quad (\gamma_i - \mbox{the set of all first class constraints}).
\end{equation}
$G$ induces the change
\begin{equation}\tag{13}
\delta F = \left\lbrace F, G\right\rbrace^*
\end{equation}
in any quantity $F$.  In ref. [5] it is noted that the extended action
\begin{equation}\tag{14}
S_E = \int dt \left[ \dot{q}_i p_i - H_c (q_j, p_j; Q_j) - u_i I\!\!P_i - U_i \Phi_i (q_j, p_j)\right]
\end{equation}
is left unaltered under changes induced by $G$ of the form of eq. (13) provided
\begin{equation}\tag{15}
\frac{D\mu_i}{Dt}+ \left\lbrace G, H_c + v_i \gamma_i\right\rbrace^* - \delta v_i \gamma_i = 0.
\end{equation}
($\frac{D}{Dt}$ contains the explicit time derivative $\frac{\partial}{\partial t}$ as well as implicit time derivative through dependence on the coefficients $v_i = (u_i, U_i)$ occurring in eq. (14).)  Upon choosing a ``gauge''
\begin{equation}\tag{16}
U_i = \delta U_i = 0
\end{equation}
in eq. (14), $S_E$ reduces to $S_T$, the total action, whose dynamical content is that of the initial classical action
\begin{equation}\tag{17}
S_c = \int dt \mathcal{L}.
\end{equation}
In this case, we can solve eq. (15) for $\lambda_i$ in terms of $\Lambda_i$ in eq. (12); we find that 
\begin{equation}\tag{18}
G = (- \dot{\Lambda}_i + c_{ijk} \Lambda_j Q_k) I\!\!P_i + \Lambda_i \Phi_i (q_j, p_j).
\end{equation}
If $\Lambda_i^A$ is the gauge function associated with generator $G_A$, then we find , using eq. (11), that
\begin{equation}\tag{19}
\left\lbrace G_A, G_B \right\rbrace^* = G_C
\end{equation}
where
\begin{equation}\tag{20}
\Lambda_i^C = c_{ijk} \Lambda_j^A \Lambda_k^B .
\end{equation}
The canonical Hamiltonian of eq. (6) which follows from eq. (1) is of the form of eq. (9).  We thus see that the gauge algebra for the model of eq. (1) closes without the need to invoke the equations of motion or to introduce auxiliary fields, as was done in ref. [1], provided we use gauge transformations generated by $G$ of the form of eq. (18).

The explicit form of $G$ that follows from eqs. (6, 9, 18) is
\begin{align}
G &= A^i \Phi_{1i} + B^i \Phi_{2i} + \overline{\Psi}C\nonumber \\
& -\left( \dot{A}_i + \epsilon_{ijk} A^j \omega^k + \frac{1}{2} \epsilon_{ijk} B^j b^k + \frac{i}{4} \overline{C} \gamma_i \psi\right) p^i\nonumber \\
& - \frac{1}{2}\left( \dot{B}_i - 2K^2\epsilon_{ijk} A^j b^k + \epsilon_{ijk}B^j \omega^k + \frac{iK}{2}\overline{C}\gamma_i \psi\right)I\!\!P^i\nonumber\\
&-\frac{1}{2}\overline{\pi}\left(\dot{C} + \frac{iK}{2} b^j \gamma_j C + \frac{i}{2} \omega^j\gamma_j C - iK\gamma^i A_j \psi - \frac{i}{2}\gamma^j B_j \psi\right).\tag{21}
\end{align}
Eqs. (13, 21) lead to the gauge transformations
\begin{equation}\tag{22a}
\delta b_{i\mu} = - \left[ \partial_\mu A_i - \epsilon_{ijk} \left( \omega_\mu^j A^k - \frac{1}{2} b_\mu^j B^k\right) + \frac{i}{4}\overline{C}\gamma_i \psi_\mu\right]
\end{equation}
\begin{equation}\tag{22b}
\delta \omega_{i\mu} = - \frac{1}{2}\left[ \partial_\mu B_i - \epsilon_{ijk} \left(2K^2 A^j b^k_\mu - B^j \omega_\mu^k \right) + \frac{iK}{2}\overline{C}\gamma_i \psi_\mu\right]
\end{equation}
\begin{equation} \tag{22c}
\delta \psi_\mu = -\frac{1}{2} \left[ \partial_\mu C + \frac{i}{2} \gamma_j\left( K b_\mu^j + \omega_\mu^j\right) C - i\gamma_j \left( KA^j + \frac{1}{2} B^j\right)\psi_\mu \right].
\end{equation}
It would be of interest to see how the local supersymmetry transformations of eq. (22) are related to the observation that supergravity is the ``square root'' of general relativity [8]. It is immediately apparent though that the transformation of eq. (22) is distinct from the supersymmetry transformations of refs. [1,2,8,12,13].  This is not surprising as the transformations of eq. (22) are a consequence of the algebra of first class constraints given in eq. (8); the usual supersymmetry transformations are a result of an algebra that is an extension of the Poincar\'e algebra. The transformations of eq. (22) are further distinguished as their algebra closes without the need for introducing auxiliary fields or invoking the equations of motion.

We note that in ref. [9] the BRST transformations associated with the $K = 0$ limit of eq. (22) are derived and that a demonstration is provided of how quantum effects can be computed using a regularization scheme that respects these local gauge symmetries.  Currently we are examining these issues for $K \neq 0$.

We also plan to apply the HTZ approach used above to find the gauge transformations in the Plebenski formulation of supergravity [10,11]. Finding such gauge transformations may be of importance in $N = 8$ supergravity because there appear to be some as yet undiscovered symmetry that would explain unexpected cancellation of divergences appearing at higher order in the perturbative loop expansion.

\section*{Appendix}
The metric $(+, -, -)$ is used with imaginary Dirac matrices $(\gamma^0, \gamma^1, \gamma^2) = (\sigma_2, i\sigma_3, i\sigma_1)$ so that
\begin{equation}\tag{A.1}
\gamma^i\gamma^j = \eta^{ij} + i\epsilon^{ijk} \gamma_k.
\end{equation}
and
\begin{equation}\tag{A.2}
\gamma^0 \gamma^i \gamma^0 = -\gamma^{iT} = \gamma^{i\dagger}.
\end{equation}
(Latin indices $i,j\ldots$ are target space; Greek indices $\mu,\nu\dots$ are space-time; early Greek $\alpha ,\beta\ldots$ are spatial.)  We use two component spinors $\psi$ satisfying that Majorana conditions
\begin{equation}\tag{A.3}
\psi = - \gamma^0 \overline{\psi}^T = \psi^* \quad (\overline{\psi} \equiv \psi^\dagger\gamma^0)
\end{equation}
and so
\begin{equation}\tag{A.4}
\overline{\psi} \chi = \overline{\chi}\psi, \quad \overline{\psi} \gamma^i \chi = - \overline{\chi} \gamma^i \psi .
\end{equation}
The Fierz identity is also useful
\begin{equation}\tag{A.5}
(\gamma^i)_{ab} (\gamma_i)_{cd} = -\frac{1}{2} (\gamma^i)_{ad} (\gamma_i)_{cb} + \frac{3}{2} \delta_{ad} \delta_{cb}.
\end{equation}
We use the left derivatives for Grassmann variable $\theta_A$ so that
\begin{equation}\tag{A.6}
\frac{d}{dt} \left(\theta_A\theta_B\right) = \dot{\theta}_A \theta_B - \dot{\theta}_B \theta_A.
\end{equation}
If $\left(q_i, p_i = \frac{\partial L}{\partial\dot{q}_i}\right)$ and $\left( \psi_i, \pi_i = \frac{\partial L}{\partial\dot{\psi}_i}\right)$ are Bosonic and Fermionic canonical variables respectively, then we define the Poisson brackets of $M$ and $N$ to be 
\begin{align}
\left\lbrace M,N \right\rbrace &= \left( M_{,q} N_{,p} - (-1)^{\epsilon_M\epsilon_N} N_{,q}M_{,p}\right) + (-1)^{\epsilon_M\epsilon_N} \left( M_{,\psi}N_{,\pi}\right.\nonumber \\
&\left. \qquad\quad + (-1)^{(\epsilon_M+1)(\epsilon_N+1)} N_{,\psi} M_{,\pi}\right)\tag{A.7}\\
& = -(-1)^{\epsilon_M\epsilon_N} \left\lbrace N,M\right\rbrace\nonumber
\end{align}
where $\epsilon_X = 0$ if $X$ is Bosonic and $\epsilon_X = 1$ if $X$ is Fermionic.

\section*{Acknowledgements}
Roger Macleod made a helpful suggestion.

\end{document}